\documentclass[a4paper,12pt]{article}
\usepackage[T1]{fontenc}
\usepackage[utf8]{inputenc}
\usepackage{cite}

\usepackage{fullpage}
\usepackage[paperheight=29.7cm, paperwidth=21.0cm,left=2.54cm,right=2.54cm,top=2.54cm,bottom=2.54cm]{geometry}

\usepackage{booktabs, makecell}

\usepackage{graphicx} 
\usepackage{amsmath}
\usepackage{amssymb}
\usepackage{mathtools}
\usepackage{wasysym}

\usepackage[svgnames]{xcolor}
\usepackage[hidelinks]{hyperref}

\date{\vspace{-9ex}}

\title{{\bf Hydrodynamics of vortex memory system driven by edge-current and boundary magnetization}}

\begin{document}
\maketitle

\begin{center}
\author{Rabiu Musah\textsuperscript{1}}

\vspace{1.25\baselineskip}

\textsuperscript{1} Department of Applied Physics, University for Development Studies, Ghana.\\
\vspace{1.0\baselineskip}
\date{April 27, 2024}
\end{center}

\vspace{1.25\baselineskip}
\textbf{Abstract:}
In this study, a magnetohydrodynamic model is developed to study the dynamics of vortices driven by edge-current. Two modeled equations for fluid and magnetic field variables are each transformed into diffusion equation for vorticity and poisson equation for stream function. A numerical solution method is designed using a simplified Lattice Boltzmann method (LBM). The LBM-D2Q5 scheme is utilized to obtain the numerical solutions for the fluid and magnetic field variables. Understanding the hydrodynamic behavior of systems employed in vortex-based memory systems is crucial for reliability and performance optimization. Based on this motivation, the effect of applied edge-current on the hydrodynamic and magnetic vortex configurations are analyzed through numerical simulations. The impact of the boundary magnetization is also conducted, by varying the strength of the magnetic field at the bottom boundary. The obtained graphical results provide some insights into the design and operation of vortex-based memory systems for next-generation data storage applications.

\vspace{1\baselineskip}

\vspace{1\baselineskip}
\textbf{Corresponding Author/Email:} R. Musah/mrabiu@uds.edu.gh

\section{Introduction}
Vortex memory systems are promising candidates for next-generation information storage due to their potential for high-density data encoding and low energy consumption. Understanding the hydrodynamic behavior of such systems is crucial for optimizing their performance and reliability.

Many approaches are used for the generation of vortices in confied domains. The most widely emplyed method of generating vortices in confined volumes is the lid-driven cavity \cite{FarahbakhshI2009,MenduSS2013,GhorbaniMA2014,ZhuJ2017,JayPN2022}. The Lorentz body force on vortices in driven cavity flow was studied. The results show that the Stuart number and penetration depth of the Lorentz force significantly affect the vortices in the flow field, leading to the formation of two new quaternary vortices at the bottom corners and the breakdown of the primary vortex into two tertiary vortices \cite{FarahbakhshI2009}. It is aso possible to create new vortices by modulating the Reynolds number towards higher values, beyond 100 \cite{ZhuJ2017}.

Vortices can also be pproduced by driven edge-currents or domain walls \cite{MöllerM2010,LuG2020}. In the context of few-nm tracking of current-driven magnetic vortex orbits, the absolute timing of the vortex gyration with respect to the driving current was studied \cite{MöllerM2010}. The proposed effect is opposite to spin injection-induced vorticity flow discussed in Ref. \cite{HeJ2006}. The authors conducted experimental investigation to observe a steady vorticity flow in the system which drives the spin Hall current. A critical current for the domain wall transformation from the vortex wall to the transverse wall was observed due to current-driven vortex domain wall dynamics \cite{HeJ2006}.

In recent times, vortices have been deemed as potential candidates for storing information. This is because, non-volatility and perfect reproducibility may be inherent in such devices. Abrikosov vortex-based random access memory cell was proposed, where a single vortex is used as an information bit \cite{TGolod2015}. The Abrikosov vortex-based random cells were characterized by an infinite magnetoresistance between '0' and '1' states, a short access time, a scalability to nanometer sizes, and an extremely low write energy. Quite recently, a stydy using computational fluid dynamics simulations revealed that vortices can be employed for conducting certain types of computation. The authors' results showed that optimal computational performance can be achieved near some critical Reynolds number, where flows exhibit a twin vortex configuration \cite{KGoto2021}. Recently, the concept of a ferroelectric vortex memory device enabling a two-bit memory element and easier vortex chirality switching was envisaged \cite{XiongH2022}. The findings suggest an alternative possibility for ferroelectric vortex-based memory devices. The study also demonstrates the deterministic switching of vortex chirality and the possibility of developing a new kind of vortex-based memory device using moderate electric field.

In this study, we investigate the hydrodynamic and magnetic vortex production due to driven edge-current and bottom boundary magnetization. This is expected to be a novel approach that harnesses the unique properties of edge currents to manipulate and store information in vortex configurations. The computational simulations are reached through Lattice Boltzman Method (LBM) within the D2Q5 scheme.

\section{Hydrodyamic model}
\subsection{Formulation of fluid and magnetic induction models}
We begin by considering the hydrodynamic equations governing the behaviour of electronic fluid in a square cavity. The Navier-Stokes equations coupled with the Maxwell equation for induced magnetic field provide a comprehensive magnetohydrodynamic framework for describing the dynamics of vortices subject to external magnetic forces.

The normalized governing equations for the fluid is described by the Navier-Stokes equation \cite{Bozkaya2011,Yu2017};
\begin{eqnarray}
\vec{\nabla}\cdot\vec{u} &=& 0,\label{eq1a}\\
\frac{\partial \vec{u}}{\partial t} + \vec{u}\cdot\vec{\nabla}\vec{u} &=& -\vec{\nabla}p + Re^{-1}\Delta\vec{u} + Al\vec{F}_{ext}.\label{eq1b}
\end{eqnarray}
The first equation represents the continuity equation, which ensures mass conservation and the second equation is momentum equation. Here, $\vec{u}$ is the two-dimensional (2D) velocity vector and $p$ is the pressure. Moreover, the Reynold's number is $Re=u_0\ell/\nu$, with $u_0$ as the reference flow velocity, $\ell$ is the characteristic size of the system, and $\nu$ is kinematic viscosity of the conducting fluid. The  external applied force is defined as $\vec{F}_{ext} = \vec{j}\times(\vec{b}+\vec{b}_0)$. Where $\vec{b}$ is the magnetic induction vector and $\vec{b}_0$ is an inplane applied field. $Al=b_0/(\mu\rho u_0^2)$ is the alfe'n number.

The governing equations for the magnetic induction are also presented in the non-dimensional form as \cite{Bozkaya2011,Yu2017};
\begin{eqnarray}
\vec{\nabla}\cdot\vec{b} &=& 0,\label{eq2a}\\
\frac{\partial \vec{b}}{\partial t} + \vec{\nabla}\times\vec{u}\times\vec{b} &=& Re_m^{-1}\Delta\vec{b}.\label{eq2b}
\end{eqnarray}
Where $Re_m = u_0\ell/\eta$ is the magnetic Reynold's number. The greek letter, $\eta$ can be viewed as the 'magnetic viscosity' and expressed as $\eta=(\mu\sigma)^{-1}$, with $\mu$ and $\sigma$ as the magnetic permeability and electrical conductivity, respectively.

\subsection{Formulation of stream function and vorticity}
To study the hydrodynamic vortex behaviour, the fluid vorticity, $\omega = \vec{\nabla} \times \vec{u}$ is introduced. The continuity equation is identically stisfied by employing the stream function, $\psi$ as; $u_x = \partial\psi/\partial y$ and $u_y = -\partial\psi/\partial x$. Thus, the vorticty-stream function formulation for equations~\ref{eq1a}-\ref{eq1b} become;
\begin{eqnarray}
\Delta\psi &=& -\omega,\label{eq3a}\\
\frac{\partial \omega}{\partial t} + \vec{u}\cdot\vec{\nabla}\omega &=& Re^{-1}\Delta\omega + S_u.\label{eq3b}
\end{eqnarray}

Where $Re=u_0\ell/\nu$ is the Reynold's number. Similarly, the 'magnetic vorticity', $Re_mj = \vec{\nabla} \times \vec{b}$ is utilized in equations~\ref{eq2a}-\ref{eq2b}. The quantity $j$ is the charge flow rate in the fluid. The $\vec{\nabla}\cdot\vec{b}=0$ condiion is identically stisfied by introducing the magnetic 'stream function', $\Lambda$ as; $b_x = \partial\Lambda/\partial y$ and $b_y = -\partial\Lambda/\partial x$. Thus, the vorticty-stream function formulation for the magnetic field equations becomes;
\begin{eqnarray}
\Delta\Lambda &=& -j,\label{eq4a}\\
\frac{\partial j}{\partial t} + \vec{u}\cdot\vec{\nabla}j &=& Re_m^{-1}\Delta j + S_b.\label{eq4b}
\end{eqnarray}

Where $Re_m=u_0\ell/\eta$ is the magnetic Reynold's number. Upon careful vector analysis, one can approximate the source terms in the preceding equation to; $S_u=Al(\vec{b}\cdot\vec{\nabla})j$ and $S_j=(\vec{b}\cdot\vec{\nabla})\omega$. Where the Alfeven number is expressed as $Al=b_0^2/(\rho\mu u_0^2)$.

\subsection{The Lattice Boltzmann formulation}
In fact, equations~\ref{eq3a}-\ref{eq4b} are fundamental equation in physics. In particular, equations~\ref{eq3a} and \ref{eq4a} are merely poisson equations and equations~\ref{eq3b} and \ref{eq4b} are advecttion-diffusion equations. Not withstanding, these equations are simply and efficiently handled within Lattice Boltzman Method (LBM) formulation \cite{Cheng2006,Chen2007,Cheng2008}. The D2Q5 configuration with BGK approximation usually employed for the poisson-advection-diffusion type models in simulations correctly reproduce physical results \cite{Cheng2008,Krüger2017,Budinski2022}. The implemetations details for the D2Q5 LBM scheme can be obtained in Ref. \cite{Mohamad2019}.

The evolution equation during a characteristic time step, $\Delta t$ for a single particle relaxation time, $\tau$, the Lattice Boltzman equation reads as:
\begin{equation}
q_{\alpha}(\vec{x}+\vec{e}_{\alpha}\Delta t, t+\Delta t) = q_{\alpha}(\vec{x},t) - \frac{q_{\alpha}(\vec{x},t) - q_{\alpha}^{eq}(\vec{x},t)}{\tau}\Delta t + \omega_{\alpha} S_{\psi}\Delta t.
\end{equation}
Where $\vec{x}$ is a 2D position vector, $t$ is the time, $q_{\alpha}(\vec{x},t)$ is the distribution function along the direction, $\alpha$ with velocity, $e_{\alpha}$ and weighing factor, $w_{\alpha}$. $q_{\alpha}^{eq}$ is the equilibrium distribution function, $S_{\psi}=-\omega$ is the source term for the stream function equation. Notice that, the source term for the vorticity equation is zero. The distribusion fuunction, $q_{\alpha}(\vec{x},t)$ is replaced with the particle distribusion function, $g_{\alpha}(\vec{x},t)$ for the stream function equation and $f_{\alpha}(\vec{x},t)$ for the vorticity transport equation. The respective equilibrium distribution functions for the D2Q5 scheme are;
\begin{eqnarray}
g_{\alpha}^{eq}(\vec{x},t) &=& w_{\alpha}\psi,\\
f_{\alpha}^{eq}(\vec{x},t) &=& w_{\alpha}\omega\left(1 + \frac{\vec{e}_{\alpha}\cdot \vec{u}}{c_s^2}\right).
\end{eqnarray}
The discrete velocity direction is expressed as;
\begin{equation}
e_{\alpha} = 
\begin{cases}
(0,0);\quad &i=0,\\
\left(\text{cos}\big(\frac{(i-1)\pi}{2}\big),\text{sin}\big(\frac{(i-1)\pi}{2}\big)\right)c;\quad &i=1,2,3,4.
\end{cases}
\end{equation}
The particle speed $c=\Delta x\, (y)/\Delta t$ and the associated weight coefficients are $w_{0} = 2/6$ and $w_{1,2,3,4} = 1/6$. The speed of sound is related to the particle speed as, $c_s = c/\sqrt{3}$. The maccroscopic variables; vorticty, $\omega$ and stream function, $\psi$ are recovered from the microscopic distribusion functions; $g_{\alpha}$, $f_{\alpha}$ as:
\begin{equation}
\psi = \sum_{\alpha}g_{\alpha},\quad \mbox{and}\quad\omega = \sum_{\alpha} f_{\alpha}.
\end{equation}

Similarly, following the above lines of arguments for the $j$-$\Lambda$ equations, the macroscopic variables, within the LBM-D2Q5 scheme, can be obtained as:
\begin{equation}
\Lambda = \sum_{\alpha}g^b_{\alpha},\quad \mbox{and}\quad j = \sum_{\alpha} f^b_{\alpha}.
\end{equation}
Where $g^b_{\alpha}$, $f^b_{\alpha}$ are the microscopic distribusion functions for the magnetic field variables.

\subsection{Boundary conditions}
For the hydrodynamic varibles, the no-slip and the so-called zero-flux boundary conditions are imposed on the flow velocity on all walls as:
\begin{equation}
\hat{n}\cdot\vec{u} = 0;\qquad \hat{n}\times\vec{u} = 0.
\end{equation}

For the induced magnetic field, an insulating boundary condition is applied to both the left and the right walls as:
\begin{equation}
\hat{n}\cdot\vec{b} = 0.
\end{equation}
However, separate boundary conditions are, employed at the bottom and top walls. At the bottom wall, where magnetization of the boundary is applied, the magnetic field boundary condition is:
\begin{equation}
b_x = b_{wall}; \qquad b_y = 0.
\end{equation}
At the top wall, where edge current is driven along the boundary, the magnetic field boundary condition is:
\begin{equation}
\frac{\partial b_y}{\partial x} - \frac{\partial b_x}{\partial y} = j_{wall}.
\end{equation}

\section{Numerical results and discussion}
Due to computational constraints, in this research a structured grid of $32\times32$ was used for numerical simulation of the problem. The LBM algorithm was simulated for about $500\times\Delta t$ seconds. Using the averaged normalized root mean squared error, for the $\omega$:
\begin{equation}
RMSE_{\omega} = \frac{\sqrt{\sum_{j=1}^N \Big[\omega_j(t+\Delta t)-\omega_j(t)\Big]^2}}{\sqrt{\sum_{j=1}^N\Big[\omega_j(t)\Big]^2}} < 10^{-5},
\end{equation}
the convergene of the silmulation is quite good, as displayed in Table~\ref{tab1} for the vorticity and stream function. Similar convergence pattern holds for the current and magnetic stream function.
\unskip

\begin{table}[!h] 
\caption{Error convergence for the numerical simulations, for $j_w=0.8$ and $b_w=0.8$.\label{tab1}}
\centering
\begin{tabular}{ccccccccccc}
\toprule [1.5pt]
\textbf{Sim. step}	&&&&& \textbf{error$_{\omega}$}	&&&&& \textbf{error$_{\psi}$}\\
\midrule
step 001	&&&&& 1.0			         &&&&& 1.0\\
step 100	&&&&& 1.11$\times10^{-4}$  &&&&& 7.21$\times10^{-5}$ \\
step 200	&&&&& 2.95$\times10^{-5}$  &&&&& 1.39$\times10^{-5}$ \\
step 300	&&&&& 4.26$\times10^{-5}$  &&&&& 8.80$\times10^{-6}$ \\
step 400	&&&&& 4.09$\times10^{-5}$  &&&&& 7.67$\times10^{-6}$ \\
step 500	&&&&& 3.10$\times10^{-5}$  &&&&& 4.71$\times10^{-6}$ \\
\bottomrule [1.5pt]
\end{tabular}
\end{table}
\unskip

Using the numerical simulations based on the modelled equations, we investigate the behavior of vortices in confined conducting fluids subjected to current flow the top edge and bottom boundary magnetization. Understanding the hydrodynamic behavior of systems employed in vortex-based memory systems is crucial for reliability and performance optimization. Thus, effects of various parameters such as; applied edge-current, $j_w$ and strength of bottom boundary magnetization, $b_w$ on the vortices behaviour are analyzed.

\begin{figure}[!h]
\centering
\includegraphics[width=14.5cm, height=8cm]{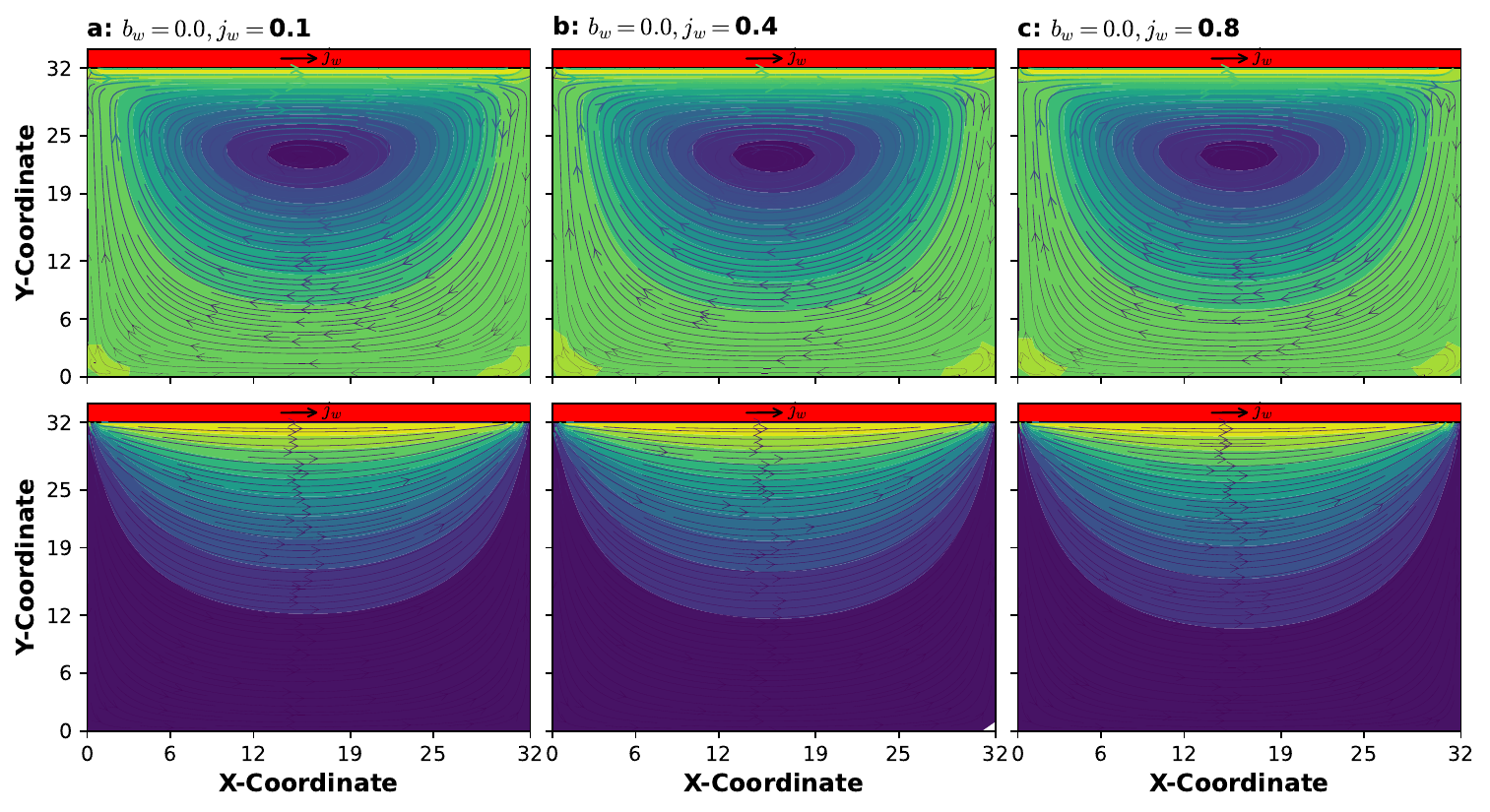}
\caption{{\bf (top)} Stream lines of $\psi$ superimposed on flow velocity fields; $u_x$ and $u_y$ contours at $b_w=0.0$. {\bf (bottom)} Stream lines of $\Lambda$ superimposed on induced magnetic fields; $b_x$ and $b_y$ contours: (\textbf{a}) $j_w=0.1$. (\textbf{b}) $j_w=0.4$. (\textbf{c}) $j_w=0.8$.\label{fig1}}
\end{figure}

In Fig.~\ref{fig1}, the impact of varying magniudes of edge-current is observed at zero bottom boundary magnetization. The top set of grapths are the stream lines, $\psi$ superimposed on flow velocity fields; $u_x$ and $u_y$ contours. As seen clearly, the usual partten of vortex configuration in a typical lid-driven cavity set-up is produced \cite{MenduSS2013,GhorbaniMA2014,ZhuJ2017}. However, the stream lines are somewhat flat near the bottom boundary that is opposite to the edge-current boundary. In this case, the varying edge-current seem to have little effect on the hydrodynamic vortex configuration. The bottom set of grapths are the magnetic stream lines, $\Lambda$ superimposed on the induced magnetic fields; $b_x$ and $b_y$ contours. Similarly, the effect of the edge-current has negligible impact on the formation of magnetic vortices. However, the moving edge-current enhances the induced magnetic field closed to the edge of the top boundary.
\unskip

\begin{figure}[!h]
\centering
\includegraphics[width=14.5cm, height=8cm]{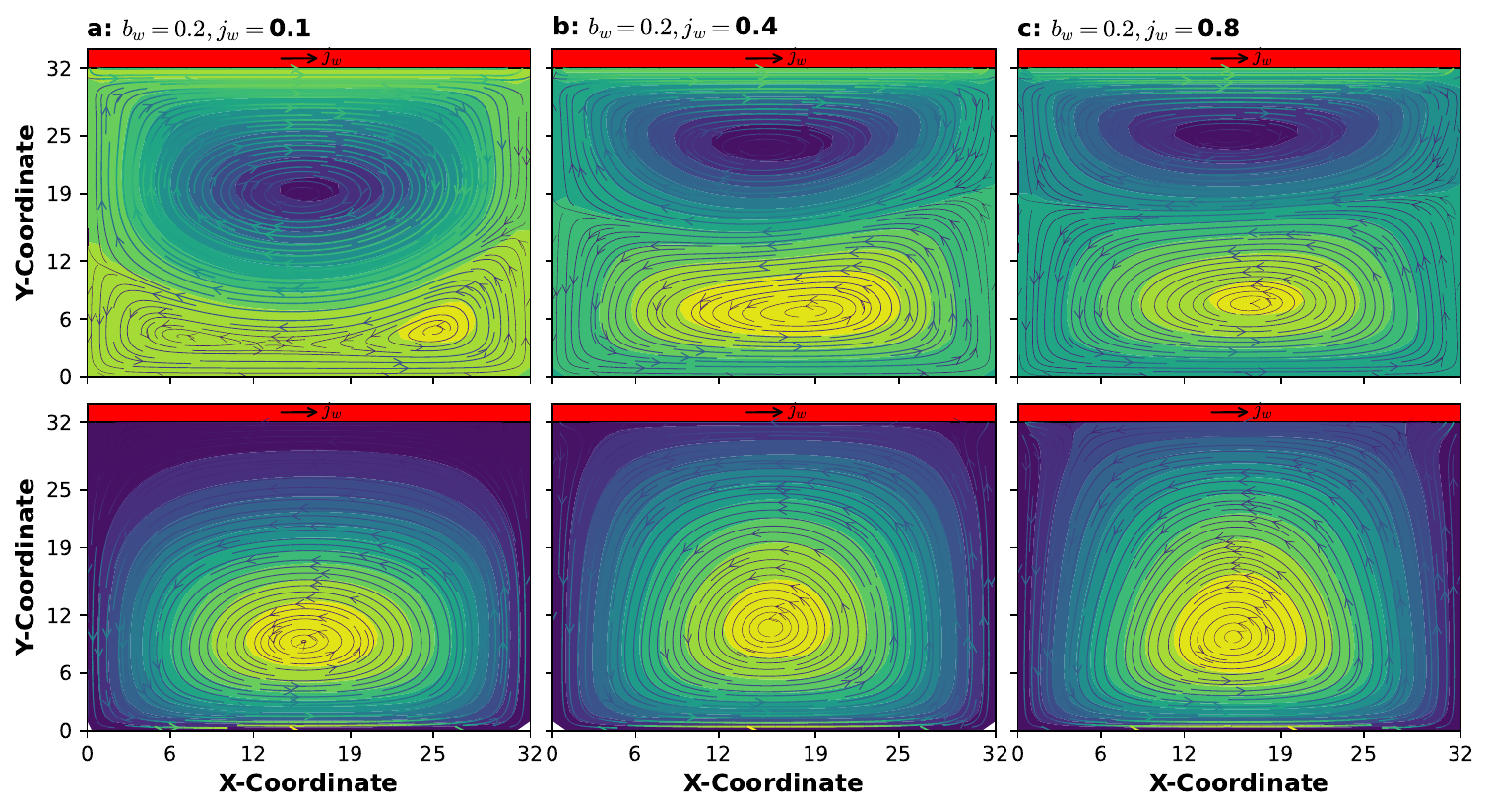}
\caption{{\bf (top)} Stream lines of $\psi$ superimposed on flow velocity fields; $u_x$ and $u_y$ contours at $b_w=0.2$. {\bf (bottom)} Stream lines of $\Lambda$ superimposed on induced magnetic fields; $b_x$ and $b_y$ contours: (\textbf{a}) $j_w=0.1$. (\textbf{b}) $j_w=0.4$. (\textbf{c}) $j_w=0.8$. \label{fig2}}
\end{figure}

In Fig.~\ref{fig2}, a startk difference occurs at varying magniudes of edge-current when the bottom boundary magnetization is different from zero, i.e $b_w=0.2$. As seen clearly in the top set of graphs, a secondary vortex generation begin to appear. This become manifest at higher edge-current values. Magnetic vortex also appear in the bottom set of graphs. This is due to the fact that the magnetized bottom boundary thickens the magnetic boundary layer. This in turn causes opposing flow to the original flow pattern. Thus, flows near the bottom boundary generates secondary vortex in the fluid domain, but with opposite chirality. As the edge-current is increased, the induced magnetic field is enhanced due to the current flow at the top boundary, which interacts positively with the magnetized field to form a pronounced effect. The vortex ceter drifts very slowly away from the magnetized boundary to the center of the bulk. In fact, the emerged secondary vortex in the fluid domain, in addition to the primary one can serve as a storage for information, say classical bits; '0' and '1'. A similar concept based on experimental investigation was recently proposed in Ref. \cite{XiongH2022}.
\unskip

\section{Conclusions}
The study sheds light on the hydrodynamics of vortex behaviour driven by edge-current. In particular, the simulations reveal the intricate interplay between hydrodynamic forces and magnetic interactions resulting in the dynamics of hydrodynamic and magnetic vortices. It is observed that edge-current can only induce hydrodynamic vortices. However, the presense of the magnetized bottom boundary creates a significant vortex behaviour for both hydrodynamic and magnetic domains, leading to potential information memory encoding and possible manipulation. The stability of both primary and the seodary vortices under different edge-current conditions provides further insights into operational regimes in the design of reliable memory storage systems. Future work may focus on experimental validations, controlling the chirality of the vortices and exploring novel material designs for enhanced vortex-based memory technologies.

\section*{Data availability statement}
The author confirm that the data supporting the findings of this study are available within the article.

\section*{Disclosure statement}
The author is not aware of any competing interests, and so has no competing interests to declare.

\end{document}